\pdfoutput=1 
\documentclass[12pt,aps,showpacs]{revtex4}%
\usepackage{graphicx}
\usepackage{amsmath}
\usepackage{amssymb}
\usepackage{amsfonts}%
\providecommand{\U}[1]{\protect\rule{.1in}{.1in}}
\begin{document}
\title{Relativistic Bessel Cylinders}
\author{J.P. Krisch and E.N. Glass}
\affiliation{Department of Physics, University of Michigan,\ Ann Arbor, MI }
\date{1 November 2014}

\begin{abstract}
A set of cylindrical solutions to Einstein's field equations for power law
densities is described.\ The solutions have a Bessel function contribution to
the metric.\ For matter cylinders regular on axis, the first two solutions are
the constant density Gott-Hiscock string and a cylinder with a metric Airy
function. All members of this family have the Vilenkin limit to their mass per
length. Some examples of Bessel shells and Bessel motion are given.

\end{abstract}

\pacs{04.20.-q, 04.20.Jb}
\maketitle

\section{Introduction}

Exact solutions in cylindrical spacetimes have found a wide variety of
applications in general relativity \cite{SKM+03}. The 1917 solutions of Weyl
and Levi-Civita are used for cylindrical vacuums \cite{BM91,WSS97,L-B14}, and
there are a wide range of cylindrical matter solutions.\ Krasinski
\cite{Kra93} has reviewed many of the cylindrical solutions with perfect and
imperfect fluid sources. Other fluid descriptions include the 1937 van Stockum
rotating dust solution \cite{v-Sto37}, constant density string solutions
\cite{His85, Got85}, cylinders with a cosmological vacuum \cite{Lin86},\ and
global solutions for static cylinders \cite{BLS+04}.\ There is interest in
cylindrical shells \cite{Sta84, BZ02}, both static and rotating \cite{KLB11}%
.\ Solutions have been combined to create van Stockum cylinders with
Gott-Hiscock cores \cite{Kri03} or matter cylinders with vacuum cores
\cite{SY12}. While infinite cylinders are not physically realized, cylindrical
structures can be used to investigate some gravitational models \cite{BK13}
and can be physically relevant for systems with cylindrical waves
\cite{Bic00}, or cosmic strings \cite{CV87, MMT+10} and have recently become
of interest \cite{BKL13, Ric13,BS14} in wormhole applications.\ 

In this paper we examine the static cylindrical field equations for the
following metric
\begin{equation}
ds^{2}=-dt^{2}+dr^{2}+F^{2}d\varphi^{2}+dz^{2},\text{ \ \ }F=F(r).
\label{cyl-met1}%
\end{equation}
We assume power law matter densities,\ and it emerges that for such densities
the field equations can be cast into a form which has solutions obeying a
Bessel\ equation. The solutions are characterized by the Bessel index or the
power law exponent.\ For integer power laws, two members of the matter filled
cylinder family are (a) the Gott-Hiscock \cite{His85, Got85} constant density
solution and (b) a solution with linear density and metric function $F(r)$ as
a combination of Airy functions. All interior solutions are matched to an
exterior vacuum Levi-Civita metric with angular deficit factor $"\alpha"$. \
\begin{equation}
ds^{2}=-dt^{2}+d\rho^{2}+dz^{2}+\alpha^{2}\rho^{2}d\varphi^{2}
\label{exterior-cyl-met}%
\end{equation}
\ Bessel shells, matched to a central Levi-Civita vacuum are also discussed
and examples of Bessel motions are given. In the next section we develop the
field equations deriving from metric (\ref{cyl-met1}), and give several
examples of the solutions for both full matter cylinders and shells.\ Among
the recurring questions about cylindrical solutions is the limit of mass per
length \cite{FG88,BLS+04}. We show that the linear mass density of the
solutions have the Vilenkin limit \cite{Vil81} and compare the possible
cylinder size limits by using the zeros of the angular deficit factor
$"\alpha".$ Several ways of adding Bessel functions to the matter motion are
discussed in the third part of the paper.

\section{The Solution Family}

\subsection*{Field Equations}

For the metric of\ Eq.(\ref{cyl-met1}), the interior field equations are (with
primes denoting $\partial/\partial r$ and units such that $G=c=1$)%
\begin{equation}
G_{tt}=-G_{zz}=-\frac{F^{\prime\prime}}{F}=8\pi\varepsilon. \label{field-eqns}%
\end{equation}
The desired Bessel form \cite{PZ95, AS72} is%
\begin{equation}
\frac{F^{\prime\prime}}{F}+\frac{a}{r}\frac{F^{\prime}}{F}+br^{n-2}=0.
\end{equation}
For simplicity, we choose $a=0$. The Bessel equation is
\begin{equation}
\frac{F^{\prime\prime}}{F}+br^{n-2}=0.
\end{equation}
where the density in Eq.(\ref{field-eqns}) was chosen as $8\pi\varepsilon
=br^{n-2}$, so that the field equation is a Bessel equation. The energy
density is either constant or zero on axis for $n\geq2.$ The $n=2$ constant
density cylinder will be used as a reference solution.\ Defining the $n=2$
density as $\varepsilon_{0}$, the power law energy density is
\begin{equation}
\varepsilon=\varepsilon_{0}(r/r_{0})^{n-2},
\end{equation}
with $\ b=8\pi\varepsilon_{0}/r_{0}^{n-2}.$\ The Bessel solution for the
metric function \cite{PZ95, AS72} is%
\begin{equation}
F(r)=\sqrt{r}\left[  A_{0}J_{1/n}(\frac{2\sqrt{b}\text{ }r^{n/2}}{n}%
)+B_{0}Y_{1/n}(\frac{2\sqrt{b}\text{ }r^{n/2}}{n})\right]  .
\end{equation}

The Kretchmann scalar for metric (\ref{cyl-met1}) is%
\begin{equation}
R^{\alpha\beta\mu\nu}R_{\alpha\beta\mu\nu}=(2\frac{F^{\prime\prime}}{F}%
)^{2}=4b^{2}r^{2(n-2)}. \label{Riem-sq}%
\end{equation}
When $n<2$ the energy density and Kretchmann scalar are singular on the axis.
For $n=2$ the curvature is constant.

Bessel functions are most easily described with a coordinate argument
$J_{1/n}(x)$. For the Bessel cylinders $x:=C(n)r^{n/2},$ with $C(n):=2b^{1/2}%
/n$ and%
\begin{equation}
F=\left[  \frac{x}{C(n)}\right]  ^{1/n}[A_{0}J_{1/n}(x)+B_{0}Y_{1/n}(x)].
\end{equation}
Herrera and Di Prisco \cite{HDiP12} have introduced a classification system
for relativistic cylinders with a general metric $g^{\text{H-DiP}}=[-A^{2},$
$A^{2},$ $B^{2},$ $C^{2}]$.\ For static anisotropic matter distributions the
solution depends on a set of three functions, $(Y_{k},Y_{s},E_{k}),$ related
to the metric functions $(A,B,C)$.\ $Y_{k}$ and $Y_{s}$ depend on metric
functions $A$ and $B$ which are zero for the Bessel cylinders. The structure
scalar, $E_{k}=F^{\prime\prime}/2F$,$\ $is determined by the energy density
through the field equations.\ As expected, the other curvature functions
(Weyl,\ Ricci, Ricci scalar) also depend on the same field relation and are
proportional to the energy density. In the next section, we describe full
matter cylinders with $n\geq2,$ where the limited range of $n$ insures a
non-singular axial density.\ 

\section{Cylinder Properties}

\subsection*{Matter Filled Cylinders: $n\geq2$}

For the special cylinders discussed here, the full matter cylinder has energy
density and axial tension with $\varepsilon=-P_{z}.$ The normalization of
$F(r)$ depends on the choice of axial behavior.\ With $n\geq2,$ the axial
density is either constant or zero. One can require that $F(r)$ approach the
interior radial coordinate near the axis.\ $F(r)$ simplifies to the Bessel
form%
\[
F(r)=A_{0}\sqrt{r}J_{1/n}[C(n)r^{n/2}].
\]
To first order, the Bessel expansion \cite{OMS00} is%
\begin{equation}
J_{1/n}[Cr^{n/2}]\text{ }{\LARGE \sim}\text{ }\frac{\sqrt{r}}{\Gamma
(\frac{n+1}{n})}[\frac{C(n)}{2}]^{1/n}\ +\ ....
\end{equation}
This determines the constant $A_{0}$. The $F(r)$ solution is \
\begin{equation}
F(r)=\Gamma(\frac{n+1}{n})\left[  \frac{2}{C(n)}\right]  ^{1/n}\sqrt{r}%
J_{1/n}[C(n)r^{n/2}],
\end{equation}
or, in terms of the x-coordinate,%
\begin{equation}
F(x)=\Gamma(\frac{n+1}{n})\left[  \frac{2}{C(n)^{2}}\right]  ^{1/n}%
x^{1/n}J_{1/n}(x).
\end{equation}
\ The angular deficit factor and interior-exterior coordinate map follow from
the metric and extrinsic curvature match to vacuum Levi-Civita at $(r_{B}$,
$\rho_{B}):$%
\begin{align}
\alpha &  =\frac{\partial F(x_{B})}{\partial r}=\Gamma(\frac{n+1}%
{n})J_{(1-n)/n}(x_{B})2^{1/n-1}[nx_{B}^{(n-1)/n}]\\
\rho_{B}  &  =\frac{J_{1/n}(x_{B})}{J_{(1-n)/n}(x_{B})}\ \frac{2x_{B}%
^{(2-n)/n\ }}{nC(n)^{2/n}}.
\end{align}
In the "Mass Density" section we show that the Bessel cylinders, in the
$n\geq2$ parameter range, have the Vilenkin upper limit on their mass per
length, the limit occurring at $\alpha=0.$ For the simplest interiors, the
zero of $J_{(1-n)/n}$ set an upper limit on $x_{B}$ (and $r_{B})$.\ In the
following section we consider two examples in the range $n\geq2:$ $n=2\ $and
$n=3.$

\subsubsection*{Matter Filled Cylinders: n = 2, 3}

The $n=2$ solution is the Gott-Hiscock \cite{His85, Got85} string with
constant density $\varepsilon=\varepsilon_{0}$, constant curvature, and axial
tension. $n=3$ describes an Airy cylinder.\ For $n=2$, the $x$ coordinate is
linearly related to $r:$ $x=kr$, with $k=\sqrt{8\pi\varepsilon_{0}}$.\ The
metric function in terms of the density is%
\begin{equation}
F(x)=\frac{\sqrt{2}}{k}\ \Gamma(\frac{3}{2})\ x^{1/2}J_{1/2}(x).
\end{equation}
Using the relation between Bessel functions of half order and spherical Bessel
functions identifies this as the Gott-Hiscock \cite{His85, Got85} constant
density string solution.\
\begin{equation}
F_{2}(x)=\frac{\sin(x)}{k},\text{ \ \ \ }\alpha=\cos(x_{B}),\text{ \ \ \ }%
\rho_{B}=\frac{\tan(x_{B})}{k}.
\end{equation}
The $n=3$ string has a linear mass density $\varepsilon=\varepsilon
_{0}(r/r_{0})$.$\ $Here
\[
x:=\frac{2}{3}kr_{0}(r/r_{0})^{3/2}.
\]
The metric function $F,$ given in terms of the x-coordinate is
\begin{equation}
F(x)=\Gamma(\frac{4}{3})\left[  \frac{2}{C^{2}(3)}\right]  ^{1/3}%
x^{1/3}J_{1/3}(x),\text{\ \ \ }C(3)=\frac{2k}{3\sqrt{r_{0}}}.
\end{equation}
The Levi-Civita match gives the angular deficit and interior-exterior
coordinate map%
\begin{align}
\alpha &  =\frac{3}{2}\Gamma(\frac{4}{3})J_{-2/3}(x_{B})(\frac{x_{B}}%
{2})^{2/3}\label{alpha}\\
\rho_{B}  &  =(\frac{k^{2}}{r_{0}})^{1/3}\frac{J_{1/3}(x_{B})}{J_{-2/3}%
(x_{B})}(\frac{2}{3x_{B}})^{1/3}. \label{rho-b}%
\end{align}
For the $n=2$ solution, the $x$ and $r$ coordinates are linearly related.\ For
$n=3$, the Bessel functions linear in $x$ are related to Airy functions with
arguments $"\chi"$ linear in the r-coordinate.%
\begin{equation}
J_{1/3}(x)=\sqrt{\frac{3}{4\chi}}\left[  \sqrt{3}Ai(-\chi)-Bi(-\chi)\right]  ,
\end{equation}
with%
\[
x=(2/3)\chi^{3/2}=C(3)r^{3/2}.
\]
The Airy functions do not simplify the analysis but provide a function linear
in $r$.\ Their occurrence is interesting because they provide a text book
example of a non-relativistic wave packet appearing as solutions to the
Schr\"{o}dinger equation with a linear potential \cite{BB79}. (In particular,
the Schr\"{o}dinger equation written with a linear potential is an Airy
equation%
\begin{equation}
\frac{d^{2}\psi}{dz^{2}}-z\psi=0. \label{airy-eqn}%
\end{equation}
This equation has solutions $Ai(z)$ for one of the two kinds of Airy
functions.)\ Here the Airy functions are the $1-d$ envelope of the cylinder
cross section and the linear behavior is in the density. \ 

\subsection*{Mass Density}

The mass per length is calculated as an integral over density.%
\begin{equation}
M=\mu L=%
{\textstyle\int\limits_{0}^{L}}
dz%
{\textstyle\int\limits_{0}^{2\pi}}
d\phi%
{\textstyle\int\limits_{0}^{r_{B}}}
\varepsilon(r)F(r)dr,
\end{equation}
with%
\[
\mu=2\pi\varepsilon_{0}\Gamma(\frac{n+1}{n})\ [\frac{2}{C(n)}]^{1/n}%
{\textstyle\int\limits_{0}^{r_{B}}}
(\frac{r}{r_{0}})^{n-2}\sqrt{r}\ J_{1/n}[C(n)r^{n/2}]dr.
\]
Changing variables, the linear density becomes
\begin{align}
4\mu &  =n\Gamma(\frac{n+1}{n})2^{(1-n)/n}\
{\textstyle\int\limits_{0}^{x_{B}}}
x^{(n-1)/n}\ J_{1/n}(x)dx\\
4\mu &  =1-\frac{\Gamma(1/n)}{2^{1-1/n}}x^{(n-1)/n}J_{(1-n)/n}(x).
\end{align}
Comparing to Eq.(\ref{alpha}), the linear density is related to the deficit
factor.%
\begin{figure}[ptb]%
\centering
\includegraphics[
natheight=1.270000in,
natwidth=1.270000in,
height=1.3689in,
width=1.3689in
]%
{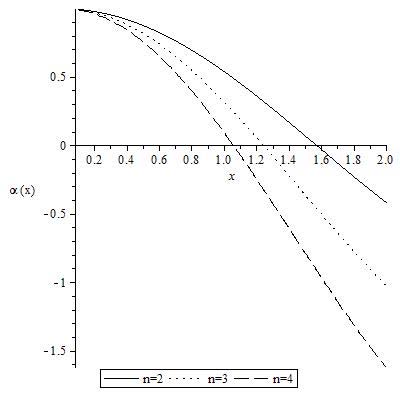}%
\caption{Angular deficit factor vs boundary position for $n=2,3,4$.}%
\end{figure}
\begin{equation}
4\mu=1-\alpha.\label{4-mu}%
\end{equation}
This relation between the linear mass density and the angular deficit factor
was originally predicted by\ Vilenkin \cite{Vil81} in a linear approximation.
Futamase and\ Garfinkle \cite{FG88} pointed out that Eq.(\ref{4-mu}) is not in
true in general, but seems to be obeyed for systems with small radial stress
\cite{Isr77}.\ There is no radial stress in these solutions and it is valid
for the Bessel cylinders. All of the $n\geq2$\ cylinders have the same upper
limit, $\mu=1/4,$ on their linear mass density.\ The cylinders differ in the
the boundary radius at which this upper limit occurs. Figure 1 shows the
angular deficit factor as a function of $x$ for several $n$ values. As $n$
increases, the zeros of the angular deficit factor and the upper limit on
$\mu$ occur at increasingly smaller values.\ The metric function $F(r)$ also
has Bessel zeros, but the first zeros of the angular deficit occur at smaller
values than the zeros of $F(r)$.\ Requiring $\alpha<1$ for positive linear
density, the angular deficit sets the upper limit on the exterior boundary. If
the string boundary is constrained to lie within the first zero, higher $n$
values can produce smaller objects.

\subsection*{Cylindrical\ Shells: n
$<$
2}

\subsubsection*{Shell Description}

Bessel shells, with no axial content, can cover the lower part of the $n$
range with a well behaved density.\ The\ general $F(x)$ is
\[
F=\left[  \frac{x}{C(n)}\right]  ^{1/n}[AJ_{1/n}(x)+BY_{1/n}(x)].
\]
The shells will match to an interior Levi-Civita vacuum solution at $x_{1}$,
with no angular deficit in the interior and to Levi-Civita solution with an
angular deficit at $x_{2}$ in the exterior.\ The matching conditions are%
\begin{align}
\lbrack\frac{x_{1}}{C(n)}]^{1/n}[A_{0}J_{1/n}(x_{1})+B_{0}Y_{1/n}(x_{1})] &
=\rho_{1}\\
\lbrack\frac{x_{2}}{C(n)}]^{1/n}[A_{0}J_{1/n}(x_{2})+B_{0}Y_{1/n}(x_{2})] &
=\alpha\rho_{2}.
\end{align}
The extrinsic curvature match provides%
\begin{align}
\lbrack A_{0}J_{(1-n)/n}(x_{1})+B_{0}Y_{(1-n)/n}(x_{1})](\frac{n}{2}%
)x_{1}^{(n-1)/n}C(n)^{1/n} &  =1\\
\lbrack A_{0}J_{(1-n)/n}(x_{2})+B_{0}Y_{(1-n)/n}(x_{2})](\frac{n}{2}%
)x_{2}^{(n-1)/n}C(n)^{1/n} &  =\alpha.
\end{align}
The three possible cases are $A_{0}$ or $B_{0}$ zero or both non zero.
$B_{0}=0$ creates shells that transition smoothly into the full matter
cylinder.\ For $B_{0}=0$ the matching conditions set the value of $A_{0}$ in
terms of the inner boundary, $x_{1}$, and the boundary parameters are%
\begin{equation}
A_{0}=(\frac{2}{n})\frac{x_{1}^{(1-n)/n}}{J_{(1-n)/n}(x_{1})C(n)^{1/n}},\text{
\ \ }\alpha=\left[  (\frac{x_{2}}{x_{1}})^{(n-1)/n}\right]  \frac
{J_{(1-n)/n}(x_{2})}{J_{(1-n)/n}(x_{1})},
\end{equation}
with boundary locations $x_{i},$ $i=1,2.$%
\begin{equation}
\rho_{i}=(\frac{2}{n})\left[  \frac{x_{i}^{(2-n)/n}}{C(n)^{2/n}}\right]
\frac{J_{1/n}(x_{i})}{J_{(1-n)/n}(x_{i})}.
\end{equation}
The $A_{0}=0$ shell is a new solution.\ For this case we have%
\begin{equation}
B_{0}=(\frac{2}{n})\frac{x_{1}^{(1-n)/n}}{Y_{(1-n)/n}(x_{1})C(n)^{1/n}},\text{
\ \ }\alpha=\left[  (\frac{x_{2}}{x_{1}})^{(n-1)/n}\right]  \frac
{Y_{(1-n)/n}(x_{2})}{Y_{(1-n)/n}(x_{1})}.
\end{equation}
The boundary locations are%
\begin{equation}
\rho_{i}=(\frac{2}{n})\left[  \frac{x_{i}^{(2-n)/n}}{C(n)^{2/n}}\right]
\frac{Y_{1/n}(x_{i})}{Y_{(1-n)/n}(x_{i})}.
\end{equation}
All choices obey the Vilenkin limit. There are differences in the position of
the boundaries and in the angular deficit.\ The $n=1$ shell \ is an
interesting example that illustrates the differences. The angular deficits and
boundary positions for $n=1$ shells are simple Bessel ratios. \
\begin{equation}
B_{0}=0:\text{ \ }\rho_{i}=\frac{2x_{i}}{C(1)^{2}}\frac{J_{1}(x_{i})}%
{J_{0}(x_{i})},\text{ \ }\alpha=\frac{J_{0}(x_{2})}{J_{0}(x_{1})},
\end{equation}%
\begin{equation}
A_{0}=0:\text{ \ }\rho_{i}=\frac{2x_{i}}{C(1)^{2}}\frac{Y_{1}(x_{i})}%
{Y_{0}(x_{i})},\text{ \ }\alpha=\frac{Y_{0}(x_{2})}{Y_{0}(x_{1})}.
\end{equation}
The interior and exterior boundary positions for a given choice of boundary
$"x"$ are shown in Figure 2 where the possible smaller core radii for the
$B_{0}=0$ shell are seen. From Figure 2, an exterior boundary value,
$x_{2}=0.4$, was selected for both shells. Figure 3 shows the variation in the
angular deficit as the interior boundary position changes. The $A_{0}=0$ shell
has much larger angular deficits for small cores (thick shells).\ As the
shells get thinner, with the inner core boundary moving away from axis, the
singular axial behavior of $Y_{1/n}$ ceases to dominate and the deficit
approaches one. \ When the two boundaries coincide, $\alpha=1,$ then the mass
per length is zero [see Eq.(\ref{4-mu})] and the entire spacetime is vacuum
Levi-Civita. \
\begin{figure}[ptb]%
\centering
\includegraphics[
natheight=1.270000in,
natwidth=1.270000in,
height=1.3689in,
width=1.3689in
]%
{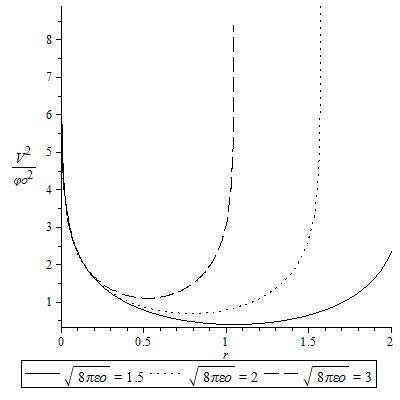}%
\caption{$V^{2}/\phi^{2}$ vs r for $n=2$and several density choices}%
\end{figure}
\begin{figure}[ptb]%
\centering
\includegraphics[
natheight=1.270000in,
natwidth=1.270000in,
height=1.3689in,
width=1.3689in
]%
{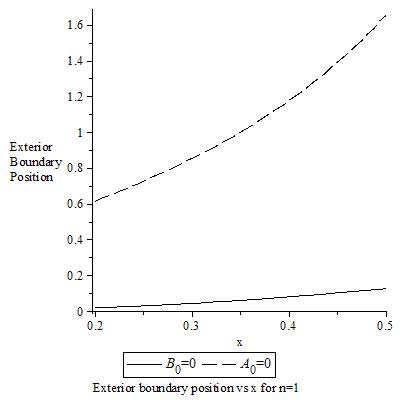}%
\caption{Boundary $\U{3c1} $ vs boundary position $x$ for $n=1.$}%
\end{figure}

\section{Describing the Motion}

\subsection*{Geodesics}

In these models, although the matter in the cylinder is static, the geodesics
contain interesting information about the cylinders and the boundary
choices.\ In the interior, the $\phi$ geodesic provides angular momentum
conservation:%
\begin{equation}
\frac{d\phi}{d\tau}F^{2}=\phi_{0}.
\end{equation}
Close to the axis, $F\sim r,$ and the equation is analogous to $\omega
r^{2}=const$.\ The radial geodesic is%
\begin{equation}
\frac{d^{2}r}{d\tau^{2}}=(\frac{F^{\prime}}{F})\phi_{0}^{2}.
\end{equation}
Converting to radial derivatives we have%
\begin{equation}
\frac{\dot{r}^{2}}{2}=\phi_{0}^{2}\ln F+const.
\end{equation}
This can be written as%
\begin{equation}
\frac{\dot{r}^{2}}{2}+V^{2}=const.
\end{equation}
where $V^{2}$ is an effective potential \cite{Sch85} in the interior given by
$V^{2}=-\phi_{0}^{2}\ln F$.%
\begin{equation}
V^{2}=-\phi_{0}^{2}\ln\left[  \Gamma(\frac{n+1}{n})[\frac{2}{C(n)^{2}}%
]^{1/n}x^{1/n}J_{1/n}(x)\right]  .
\end{equation}
For example, with $n=2,$ $k=\sqrt{8\pi\varepsilon_{0}}$, and $x=kr$, we have
$V^{2}=-\phi_{0}^{2}\ln(\frac{\sin(kr)}{k})$.\ The potential is concave upward
with a stable equilibrium at $F^{\prime}=0$. The equilibrium is at a zero of
$J_{-1.2}(kr)$, with the first zero occurring at $kr\sim1.57.$ The shape of
the potential suggests that the matter interior is stable under small radial
perturbations with boundary chosen inside the potential. \ The shape of the
potential for $n=2$ and several density choices are shown in Figure 4, which
illustrates that the choice of exterior boundary is not only dependent on the
allowed boundary ranges but also on the potential shape.%
\begin{figure}[ptb]%
\centering
\includegraphics[
natheight=1.270000in,
natwidth=1.270000in,
height=1.3689in,
width=1.3689in
]%
{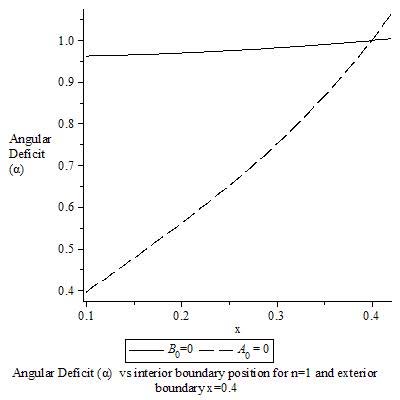}%
\caption{Angular deficit $\U{3b1} $ vs x.}%
\end{figure}

\subsection*{Motion with a Bessel Description}

Bessel functions can be added to a fluid description by the choice of velocity
profile.\ Two examples are considered.\ The first example looks at the
conditions for an irrotational motion described by\ Bessel fuctions.\ In the
second example, the geodesic description is used to add Bessel behavior to a
radial velocity.

\subsubsection*{Tangential Motion}

A 4-velocity with a rotational component $U^{i}=[U^{0},0,U^{\varphi},0]$\ has
vorticity in the z direction proportional to
\begin{equation}
\omega^{z}\sim\frac{\partial(U^{\varphi}F)}{\partial r}.
\end{equation}
An irrotational motion will have $U^{\varphi}\sim c_{1}/F(r).$ Close to the
axis, this is the $c_{1}/r$ velocity profile associated with classical
point-vortex motion. For thick shells, near the outer boundary, the velocity
profile, while still irrotational, is no longer that of a point vortex.\ The
shells discussed here have Levi-Civita vacuum interiors. They would need a
matter core to describe vortices, for example, of the Rankine type.

\subsubsection*{Radial Motion}

Radial motion with a Bessel function description can be added by considering
the geodesics of test particles.\ They involve a non-zero angular momentum
parameter so the actual motion would include some rotational component. The
radial geodesic equation can be written with an extra derivative as%
\begin{equation}
\frac{d^{2}}{dr^{2}}(\frac{\dot{r}^{2}}{2})+\frac{d}{dr}(\frac{d\ln F}{dr})=0.
\end{equation}
This is an Airy equation for $\dot{r}^{2}/2$ if the second term has the form%
\begin{equation}
\frac{d}{dr}(\frac{d\ln F}{dr})=-r(\frac{\dot{r}^{2}}{2})=-r(\phi_{0}^{2}\ln
F+const).
\end{equation}
putting Airy function into the radial velocity.\ This example changes the
metric function. Using the geodesic solution, this becomes a generating
equation for a new $F(r)$:%
\begin{align}
F(r)  &  =e^{[\lambda(r)-1]/\phi_{0}^{2}}\\
\frac{d}{dr}(\frac{d\ln F}{dr})  &  =-r(\phi_{0}^{2}\ln F+1)\\
\frac{d^{2}\lambda}{dr^{2}}+(\phi_{0}^{2})r\lambda &  =0.
\end{align}
This is a particular Bessel equation of order $1/3$. Again, there is Airy
behavior in metric function $F(r)$, but now it appears exponentially with
non-zero angular momentum parameter $\phi_{0}^{2}$.

\section{Discussion}

A family of Bessel solutions of the Einstein equations with power law density
$\varepsilon\sim r^{n-2}$, and axial stress were developed. For $n\geq2$, the
first two solutions are the $n=2$ Gott-Hiscock string and an $n=3$ Airy
cylinder.\ All cylinders have the Vilenkin upper limit on their mass per
length. The cylinders reach that limit at increasing smaller radii. The choice
of power law density was motivated by recent experiments using lasers
\cite{SC07,SBD+08} and electron beams \cite{BLL+13} that reported
observational evidence for beam caustics described by Airy functions. Airy
functions are normally associated with a linear potential in the
Schr\"{o}dinger equation.\ The Airy functions here are in the metric, and the
linear dependence occurs for the $n=3$ density. The particular $n=3$ Airy
solution discussed here is part of the Gott-Hiscock family, and there are
other related solutions containing Airy functions.\ 

Most of the Bessel cylinders described in this work are static but extensions
to stationary or time dependent solutions for metrics containing Airy
functions would also be of interest for some cylindrical beam applications.
The cylinders studied here can have a fractional Bessel index, but are
solutions of an integer derivative Bessel equation. Using the relation between
the fractional derivative of the sine \cite{OMS00} and Bessel functions
provides%
\[
J_{1/n}(\sqrt{z})=\frac{d^{(n-2)/2}}{d^{(n-2)/2}}\sin(\sqrt{z}).
\]
Links to fractional calculus provide a possible direction for future work. \

\end{document}